\documentclass{article}

\usepackage[letterpaper, total={6.5in, 9in}]{geometry}
\usepackage{graphicx}

\usepackage[backend=biber,natbib,sortcites,sorting=none,
citestyle=numeric-comp,  
bibstyle=nejm,
maxcitenames=2,mincitenames=1,maxbibnames=4,
isbn=false,url=false,doi=false
]{biblatex}

\usepackage[scaled=0.8]{dictsym}
\usepackage{booktabs}
\usepackage{amsmath}
\usepackage{rotating}
\usepackage{tablefootnote}
\usepackage{multirow}
\usepackage{adjustbox}
\usepackage{subcaption}
\usepackage{ragged2e}
\usepackage{amssymb}

\usepackage{preprint}

\newcommand{\comment}[1]{}

\newcommand*{\writecomment}[2]{\csdef{localcomment#1}{#2}}

\renewbibmacro*{finentry}{%
  \finentry
  \ifcsdef{localcomment\thefield{entrykey}}
    {\\[0.5em]\csuse{localcomment\thefield{entrykey}}}
    {}%
}

\newcommand{\emphref}{\raisebox{0.85ex}{\scalebox{0.4}{$\blacksquare$}}}
\newcommand{\emphrefbib}{\raisebox{0.3ex}{\scalebox{0.6}{$\blacksquare$}}}

\DeclareBibliographyCategory{important}
\DeclareBibliographyCategory{veryimportant}
\DeclareFieldFormat{labelnumber}{\ifcategory{veryimportant}{#1\emphref\emphref}{\ifcategory{important}{#1\emphref}{#1}}}

\newcommand{\importantref}[2]{
\addtocategory{important}{#1}
\writecomment{#1}{ #2}
}

\newcommand{\veryimportantref}[2]{
\addtocategory{veryimportant}{#1}
\writecomment{#1}{ #2}
}

\newcommand{\textdsmedical}{\mbox{\hspace{-1.5px}\protect\rotatebox[origin=c]{35}{\protect\raisebox{-1.2pt}{\small\mbox{\protect\dsmedical}}$\!$}}}

\newcommand{\textdsmedicaltiny}{\mbox{$\!$\protect\rotatebox[origin=c]{35}{\protect\raisebox{-1pt}{\scriptsize\mbox{\protect\dsmedical}}$\!$}}}

\importantref{Consejo2018IntroductionOphthalmologists}{This survey on machine learning provides a good overview of the field and is written to be accessible to ophthalmologists.}

\importantref{Beers2018High-resolutionNetworks}{This is the first example of realistic synthesized ROP fundoscopic images. Synthesized images would be an effective way to augment data sets and resident education without compromising patient privacy.}

\importantref{Brown2018AutomatedNetworks}{The i-ROP-DL system detects plus disease in infants with ROP more accurately than the majority of experts in this study. This article highlights a deep learning method with the ability to surpass physician performance.}

\importantref{Lin2018PredictionStudy}{This study predicts the development of high myopia in children up to 8 years in advance. Such prediction could potentially be used to guide atropine prophylaxis.}

\importantref{Liu2017LocalizationNetwork}{This study describes a cloud-based ML platform, CC-Cruiser, that accurately detects cataract presence, area, density, and location. Such an approach could detect cataracts in the primary care setting or serve as a complement to the pediatric ophthalmologist's evaluation.}

\veryimportantref{Redd2018EvaluationPrematurity}{The i-ROP-DL deep learning system is the first to detect specific ROP classifications, including clinically significant, type 1, and type 2 ROP. This model could potentially be a useful telemedical tool for identifying referral-warranted ROP.}

\veryimportantref{Wang2018AutomatedNetworks}{The DeepROP system for ROP detection is trained on the largest data set to date, and is the first to detect severe ROP using fundus images that include the peripheral retina. This deep learning approach demonstrates the potential benefits of fine-grained ROP classification.}

\veryimportantref{Lu2018AutomatedApplications}{This system is the first to detect strabismus remotely from digital facial images. As a telemedical application, this could help determine which children require an ophthalmology referral for strabismus.}

\veryimportantref{Vogelsang2018PotentialAcuity}{This article proposes that high initial acuity can disrupt visual development, and suggests it as an explanation of why adults with a history of congenital cataract surgery in infancy may exhibit deficient facial recognition. Their hypothesis is supported by experimental results that use convolutional neural networks to model visual development, and could be used to improve neural network training.}

\importantref{Lin2019DiagnosticTrial}{This study describes a multi-center randomized controlled trial evaluating the performance of the CC-Cruiser system for cataract diagnosis and treatment---an important step toward a real-world clinical application of AI to pediatric ophthalmology.}

\title{Artificial Intelligence for Pediatric Ophthalmology}
\author{\textit{Julia E.~Reid, MD}$^{\textdsmedical,\dagger}$
and \textit{Eric Eaton, PhD}$^\ddagger$
\thanks{$^{\textdsmedicaltiny}$Nemours / Alfred I.~duPont Hospital for Children, Division of Pediatric Ophthalmology, Wilmington, DE; $^\dagger$Thomas Jefferson University, Departments of Pediatrics and Ophthalmology, Philadelphia, PA; and $^\ddagger$University of Pennsylvania, Department of Computer and Information Science, Philadelphia, PA
\protect\\~\protect\\
Correspondence to Julia E.~Reid, MD, Division of Pediatric Ophthalmology, 1600 Rockland Road, Wilmington, DE 19803, USA. email: julia.e.reid@nemours.org}
}

\begin{document}

\twocolumn[
\maketitle

\begin{abstract} 

\noindent\textbf{Purpose of review}\\Despite the impressive results of recent artificial intelligence (AI) applications to general ophthalmology, comparatively less progress has been made toward solving problems in pediatric ophthalmology using similar techniques.  This article discusses the unique needs of pediatric ophthalmology patients and how AI techniques can address these challenges, surveys recent applications of AI to pediatric ophthalmology, and discusses future directions in the field.

\noindent\textbf{Recent findings}\\The most significant advances involve the automated detection of retinopathy of prematurity (ROP), yielding results that rival experts. Machine learning (ML) has also been successfully applied to the classification of pediatric cataracts, prediction of post-operative complications following cataract surgery, detection of strabismus and refractive error, prediction of future high myopia, and diagnosis of reading disability via eye tracking. In addition, ML techniques have been used for the study of visual development, vessel segmentation in pediatric fundus images, and ophthalmic image synthesis.

\noindent\textbf{Summary}\\ AI applications could significantly benefit clinical care for pediatric ophthalmology patients by optimizing disease detection and grading, broadening access to care, furthering scientific discovery, and improving clinical efficiency. These methods need to match or surpass physician performance in clinical trials before deployment with patients. Due to widespread use of closed-access data sets and software implementations, it is difficult to directly compare the performance of these approaches, and reproducibility is poor.  Open-access data sets and software implementations could alleviate these issues, and encourage further AI applications to pediatric ophthalmology.



\end{abstract}~\\
]
\saythanks

\section{Introduction}

The increased availability of ophthalmic data, coupled with advances in artificial intelligence (AI) and machine learning (ML), offer the potential to positively transform clinical practice.  Recent applications
of ML techniques to general ophthalmology have demonstrated the potential for automated disease diagnosis \citep{Gulshan2016DevelopmentPhotographs}, automated prescreening of primary care patients for specialist referral \citep{DeFauw2018ClinicallyDisease}, and scientific discovery \citep{Varadarajan2018DeepImages}, among others.  Acting as a complement to ophthalmologists, these and future applications have the potential to optimize patient care, reduce costs and barriers to access, limit unnecessary referrals, permit objective monitoring, and enable early disease detection.

To date, most AI applications have focused on adult ophthalmic diseases, as discussed by several reviews 
\citep{Roach2017ArtificialIntelligence,Consejo2019IntroductionOphthalmologists,Ting2018Artificial,Lee2017MachineArrived,Rahimy2018DeepOphthalmology,Caixinha2017MachineSciences,AmericanAcademyofOphthalmology2018TheOphthalmology, Du2018ApplicationOphthalmology}.  Comparatively little progress has been made in applying AI and ML techniques to pediatric ophthalmology, despite the pressing need.  In the United States, there is a shortage of pediatric ophthalmologists \citep{Estes2007TheProject} and fellowship positions continue to go unfilled \citep{Dotan2017Pediatric2000-2015}. Globally, this shortage is even more pronounced and devastating---for example, retinopathy of prematurity (ROP), now in its third epidemic, has resulted in irreversible blindness in over 50,000 premature infants due to worldwide shortages of trained specialists and other barriers to adequate care \citep{Gilbert2008RetinopathyControl,Quinn2016RetinopathyEpidemic}.

\section{Unique considerations for\\pediatric ophthalmology}
Ophthalmic disease prevalence, cause, presentation, diagnosis, and treatment all differ between adult and pediatric patients---dissimilarities that are important to consider when developing AI applications.

Common diseases in children include amblyopia, strabismus, nasolacrimal duct obstruction (NLDO), retinopathy of prematurity (ROP), and congenital eye diseases. The adult population, by contrast, is affected by cataracts, dry eye, macular degeneration, diabetic retinopathy, and glaucoma. For diseases that occur in both children and adults, the presentation, cause, and treatment often differ. Glaucoma is a good example, as the cause and presentation in congenital glaucoma patients are both unlike those in adult-onset glaucoma patients. Optimal management of glaucoma, including surgery, also differs for these two populations.

\keypoints{
\item Pediatric ophthalmology has unique aspects that must be considered when designing AI applications, including disease prevalence, cause, presentation, diagnosis, and treatment, which differ from adults. 
\item Most recent AI applications focus on ROP or congenital cataracts, although many other areas of pediatric ophthalmology could benefit from AI.
\item Reproducibility and comparability between current AI approaches is poor, and would be improved with open-access data sets and software implementations.
\item Evaluation on experimental data sets should be augmented with clinical validation prior to deployment with patients.}

Infants and children have distinct characteristics from adults that affect their ophthalmology visits. Given their developmental capabilities, there is generally less information gleaned from a single eye exam of a child, so several visits may be required to accurately diagnose or characterize that child's disease. There is also a stronger reliance on the objective exam because of the infant's or child's inability to effectively communicate. Children's short attention spans and unpredictable behavior often necessitate a quick exam that allows the physician to gain the child's trust while keeping him or her at ease. Despite this, there are portions of the clinic visit that take longer, such as restraining a child to administer dilating drops and then waiting for that child to be fully cyclopleged. Ancillary testing that requires patient cooperation may not be possible in an awake child, and eye exams under anesthesia are not uncommon. Similarly, children are typically placed under general anesthesia for eye procedures, whereas adults may require only topical or local anesthesia.  Techniques for more accurate diagnosis and disease prediction could help reduce the high cost and risk of repeated exams and surgeries under anesthesia.

Other distinguishing factors pertain to the pediatric patient's growth and development. In most children, visual development occurs from birth until age 7 or 8; eye diseases affecting children 
during this period can cause permanent vision loss due to amblyopia or reduced visual abilities. Additionally, during development, significant ocular 
growth occurs, causing changes in refractive error that complicate surgical planning for congenital cataract patients. 

Retinal imaging, too, differs for pediatric and adult patients. Factors such as children's lack of fixation and small pupils can create blur, partial occlusion, and illumination defects, all of which degrade image quality. For infants being screened for ROP, their fundus images are more variable and have more visible choroidal vessels, making classification comparatively difficult \citep{Worrall2016AutomatedNetworks}.

\newlength{\tblhangindention}
\setlength{\tblhangindention}{0.75em}
\newcommand{\tblindent}{\hspace*{\tblhangindention}}
\newcommand{\NR}{\hspace*{.5em}--}

\begin{table*}[!tb]
    \centering
    \footnotesize
    
    \renewcommand{\arraystretch}{1.098} 
    \setlength{\tabcolsep}{0.8\tabcolsep}
    \setlength{\aboverulesep}{0ex} 
    \setlength{\belowrulesep}{0ex}
    \sffamily

    \caption{Summary of ML-based techniques for pediatric ophthalmic disease detection and diagnosis}
    \label{tab:ROPsummary}

    \noindent\adjustbox{max width=\textwidth}{
    
    \begin{minipage}{1.0135\textwidth}
    \begin{tabular}{@{\kern\tabcolsep}p{1.1in}p{1.5in}p{0.45in}p{0.45in}p{0.35in}p{0.4in}p{1.5in}@{\kern\tabcolsep}} 
    \toprule
      \rowcolor{headercolor} \textbf{Approach \newline (Approx.~devel.~year)}   &  \textbf{Predicted category} &  \textbf{Sensitivity\newline (\%)} & \textbf{Specificity\newline (\%)} & \textbf {AUROC} & \textbf {Accuracy\newline (\%)}& \textbf{Method summary}\\\midrule
      
      \multicolumn{2}{l}{\textbf{Retinopathy of prematurity (ROP)}}   &  & \\
      
       \tblindent DeepROP \citep{Wang2018AutomatedNetworks} \newline \tblindent (\citeyear{Wang2018AutomatedNetworks})
      & Experimental data set 
      \newline \tblindent Presence of ROP \newline 
      \tblindent Severe (vs Mild) ROP \newline
      Clinical test 
      \newline \tblindent Presence of ROP \newline 
      \tblindent Severe (vs Mild) ROP
      & ~\newline 96.64 \newline 88.46 \newline \newline 84.91 \newline 93.33     
      & ~\newline 99.33 \newline 92.31 \newline \newline 96.90 \newline 73.63
      & ~\newline 0.995 \newline 0.951 \newline \newline \NR \newline \NR  
      & ~\newline 97.99 \newline 90.38 \newline \newline 95.55 \newline 76.42
      & Cloud-based platform. Set of fundus images $\rightarrow$ two CNNs (modified Inception-BN nets pretrained on ImageNet): one predicts presence, and the other severity\\
      
      \tblindent i-ROP-DL \citep{Redd2018EvaluationPrematurity} 
      \newline \tblindent (\citeyear{Redd2018EvaluationPrematurity})
      & Clinically significant ROP \newline \tblindent Type 1 ROP \newline \tblindent  Type 2 ROP \newline \tblindent Pre-plus disease
      & \NR                         \newline 94         \newline \NR           \newline \NR               
      & \NR                         \newline 79         \newline \NR           \newline \NR
      & 0.914                      \newline 0.960        \newline 0.867        \newline 0.910
      & \NR \newline \NR \newline \NR \newline \NR
      & Applies a linear formula to the probabilities output by  \mbox{i-ROP-DL}  (see below) to yield a severity score on a 1--9 scale\\
      
      \tblindent  MiGraph \citep{Rani2016MultiplePrematurity} 
      \newline \tblindent (2016)
      & Presence of ROP 
      & 99.4
      & 95.0
      & 0.98
      & 97.5
      & SIFT features from image patches $\rightarrow$ multiple instance learning graph-kernel SVM\\
      
      \tblindent  VesselMap \citep{Rabinowitz2007ProgressionAge} 
      \newline \tblindent (\citeyear{Rabinowitz2007ProgressionAge})
      & Severe ROP \newline \tblindent \mbox{From mean arteriole diameter} \newline \tblindent From mean venule diameter
      & ~\newline \NR \newline \NR 
      & ~\newline \NR \newline \NR 
      & ~\newline 0.93 \newline 0.87
      & ~\newline \NR \newline \NR
      & Semiautomated tool that uses classic image analysis to measure vessel diameter
      \global\rownum=1\relax  
      \\
    
      \midrule
      
      \multicolumn{2}{l}{\textbf{ROP: Plus or pre-plus disease}}   &  & &  & & \\
    
     \mbox{\tblindent  i-ROP-DL  \citep{Brown2018AutomatedNetworks}} \newline  
    \tblindent (\citeyear{Redd2018EvaluationPrematurity})
      &  Plus disease \citep{Redd2018EvaluationPrematurity} \newline Pre-plus disease \citep{Redd2018EvaluationPrematurity} 
      \newline Plus disease \citep{Brown2018AutomatedNetworks}
      \newline Pre-plus or worse disease \citep{Brown2018AutomatedNetworks}
      & \NR \newline \NR \newline 93 \newline 100
      & \NR \newline \NR  \newline 94 \newline 94
      & 0.989 \newline 0.910 \newline 0.98 \newline 0.94
      & \NR \newline \NR \newline 91.0 \newline \NR
      & CNN-output (U-net) vessel segmentations $\rightarrow$ CNN (InceptionV1 pretrained on ImageNet) to classify as normal/pre-plus/plus\\
      
    \tblindent  CNN + Bayes \citep{Worrall2016AutomatedNetworks}\newline \tblindent (\citeyear{Worrall2016AutomatedNetworks})
      & Plus disease (per image) \newline \phantom{Plus disease} (per exam)
      & 82.5 \newline 95.4 
      & 98.3 \newline 94.7 & \NR \newline \NR
      & 91.8 \newline 93.6
      & CNN (InceptionV1 pretrained on ImageNet) adapted to output the Bayesian posterior \\

      \tblindent i-ROP \citep{Ataer-Cansizoglu2015Computer-basedDiagnosis} \newline \tblindent (2015) & Plus disease  \newline Pre-plus or worse disease
      & 93 \newline 97
      & \NR \newline \NR
      & \NR \newline \NR
      & 95 \newline \NR &
      SVM with a kernel derived from a GMM of tortuosity and dilation features from manually segmented images 
      \\

     \mbox{\tblindent  Na\"ive Bayes  \citep{Bolon-Canedoa2015DealingApproach}}
      \newline \tblindent (2015)
      & \mbox{Plus/pre-plus/none (SVM-RFE)} \newline Plus disease (ReliefF)
      & \NR \newline \NR
      & \NR \newline \NR
      & \NR \newline \NR
      & 79.41 \newline 88.24
      & Na\"ive Bayes with SVM-RFE or ReliefF vessel feature selection\\
      
      \tblindent  CAIAR \citep{Shah2009SemiautomatedPrematurity}   
      \newline \tblindent (\citeyear{Wilson2008ComputerizedInfants})
      & Plus (from venule width) \newline Plus (from arteriole tortuosity) 
      & \NR \newline \NR
      & \NR \newline \NR
      & 0.909 \newline 0.920
      & \NR \newline \NR
      & Generative vessel model fit to a multi-scale representation of the retinal image\\
      
      \tblindent  ROPtool \citep{Wallace2007APrematurity}   
      \newline \tblindent (\citeyear{Wallace2007APrematurity})
      & Plus tortuosity (eye) \newline \phantom{Plus tortuosity} (quadrant) \newline Pre-plus tortuosity (quadrant)
      & 95 \newline 85 \newline 89
      & 78 \newline 77 \newline 82
      & \NR \newline 0.885 \newline 0.875
      & 87.50 \newline 80.63 \newline \NR
      & User-guided tool that traces centerlines of retinal vessels to measure tortuosity\\

      \tblindent  RISA \citep{Gelman2007PlusDiagnosis}  
      \newline \tblindent (2005)
      & \hangindent=\tblhangindention Plus disease (from arteriole and venule curvature and tortuosity, venule diameter)
      & 93.8 & 93.8 & 0.967
      & \NR
      & Logistic regression on geometric features computed for each segment of the vascular tree\\

      \tblindent  IVAN \citep{Shah2009SemiautomatedPrematurity}
      \newline \tblindent (2002)
      & Plus (from venule width)
      & \NR
      & \NR 
      & 0.909
      & \NR
      & Measures vessel width via classic image analysis\\
      
            \bottomrule
    \rowcolor{white}\multicolumn{7}{l}{Abbreviations:  AUROC -- area under the receiver operating characteristic curve; GMM -- Gaussian mixture model}\\
    \end{tabular}
    \end{minipage}
    }
\end{table*}

\begin{table*}[!tb]
    \centering
    \footnotesize
    
    \renewcommand{\arraystretch}{1.098} 
    \setlength{\tabcolsep}{0.8\tabcolsep}
    \setlength{\aboverulesep}{0ex} 
    \setlength{\belowrulesep}{0ex}
    \sffamily

    \addtocounter{table}{-1}  
    \caption{(Continued)} 

    \noindent\adjustbox{max width=\textwidth}{
    
    \begin{minipage}{1.0135\textwidth}
    \begin{tabular}{@{\kern\tabcolsep}p{1.1in}p{1.5in}p{0.45in}p{0.45in}p{0.35in}p{0.4in}p{1.5in}@{\kern\tabcolsep}} 
    \toprule
      \rowcolor{headercolor} \textbf{Approach \newline (Approx.~devel.~year)}   &  \textbf{Predicted category} &  \textbf{Sensitivity\newline (\%)} & \textbf{Specificity\newline (\%)} & \textbf {AUROC} & \textbf {Accuracy\newline (\%)}& \textbf{Method summary}\\\midrule
      
      \multicolumn{2}{l}{\textbf{Pediatric cataracts}}   &  & & & & \\

      \tblindent  Post-operative 
      \newline \tblindent complication\newline \tblindent prediction 
      \citep{Zhang2019PredictionMining} \newline \tblindent (\citeyear{Zhang2019PredictionMining})

      & 
      CLR and/or High IOP (RF) \newline \phantom{CLR and/or High IOP} (NB)\newline Central lens regrowth (RF) \newline \phantom{Central lens regrowth} (NB) \newline High IOP (RF)  \newline \phantom{High IOP} (NB)
      & 62.5 \newline 73.1 \newline 66.7 \newline 61.1 \newline 63.6 \newline 54.5
      & 76.9 \newline 66.7 \newline 72.2 \newline 68.8 \newline 71.8 \newline 69.2
      & 0.722 \newline 0.719 \newline 0.743 \newline 0.735 \newline 0.735 \newline 0.719
      & 70.0 \newline 70.0 \newline 72.0 \newline 66.0 \newline 70.0 \newline 66.0
      & Demographic and cataract severity evaluation data $\rightarrow$ class-balancing using SMOTE $\rightarrow$ random forest (RF) and na\"{i}ve Bayes (NB) classifiers\\
    
    \mbox{\tblindent  CS-ResCNN 
      \citep{Jiang2017AutomaticNetwork}}
      \newline \tblindent (2017)
      & Severe posterior\newline capsular opacification
      & ~\newline 89.66 
      & ~\newline 93.19 
      & ~\newline 0.9711
      & ~\newline 92.24
      & Slit-lamp images $\rightarrow$ automatically crop to lens $\rightarrow$ CNN (ResNet pretrained on ImageNet) with cost-sensitive loss\\

      \tblindent  CC-Cruiser 
      \citep{Long2017AnCataracts}
      \newline \tblindent (2016)
      & \RaggedRight Multi-center trial \newline \tblindent Cataract presence \citep{Lin2019DiagnosticTrial} \newline \tblindent Opacity area grading \citep{Lin2019DiagnosticTrial}\newline \tblindent Density grading \citep{Lin2019DiagnosticTrial}\newline \tblindent Location grading \citep{Lin2019DiagnosticTrial}\newline \tblindent Treatment  \citep{Lin2019DiagnosticTrial}
      \newline Experimental data set \tblindent Cataract presence \citep{Liu2017LocalizationNetwork}\newline \tblindent Area grading \citep{Liu2017LocalizationNetwork} \newline \tblindent Density grading \citep{Liu2017LocalizationNetwork} \newline \tblindent Location grading \citep{Liu2017LocalizationNetwork}
      & ~\newline 89.7 \newline 91.3 \newline 85.3 \newline 84.2 \newline 86.7
      \newline ~\newline 96.83 \newline 90.75 \newline 93.94 \newline 93.08
      & ~\newline 86.4 \newline 88.9 \newline 67.9 \newline 50.0 \newline 44.4
      \newline ~\newline 97.28 \newline 86.63 \newline 91.05 \newline 82.70
      & ~\newline \NR \newline \NR \newline \NR \newline \NR\newline \NR 
      \newline ~\newline 0.9686 \newline 0.9892 \newline 0.9743 \newline 0.9591
      & ~\newline 87.4 \newline 90.6 \newline 80.2 \newline 77.1 \newline 70.8
      \newline ~\newline 97.07 \newline 89.02 \newline 92.68 \newline 89.28
      & Cloud-based platform. Slit-lamp images $\rightarrow$ automatically crop to lens $\rightarrow$ three CNNs (AlexNets) to predict: cataract presence, severity (area, density, location), and treatment (surgery or follow-up)\\
      
      
      \midrule
      
      \multicolumn{2}{l}{\textbf{Strabismus}}   &  & & & & \\
      
      \tblindent  RF-CNN \citep{Lu2018AutomatedApplications} 
        \newline \tblindent (2018)
      & Strabismus presence
      & 93.30 
      & 96.17
      & 0.9865
      & 93.89
      & Two-stage CNN: eye regions segmented from face images via R-FCN $\rightarrow$ 11-layer CNN \\
     
      \tblindent  SVM + VGG-S \citep{Chen2018StrabismusNetworks} 
        \newline \tblindent (2017)
      & Strabismus presence
      & 94.1 
      & 96.0
      & \NR
      & 95.2
      & Eye-tracking gaze maps $\rightarrow$ CNN (VGG-S pretrained on ImageNet) features $\rightarrow$ SVM\\

      \tblindent  Pediatric Vision\newline\tblindent Screener \citep{Gramatikov2017DetectingScanning} 
        \newline \tblindent (\citeyear{Gramatikov2017DetectingScanning})
      & Central vs.~paracentral fixation \newline \tblindent Experimental evaluation \newline \tblindent Clinical evaluation
      & ~\newline 100.0 \newline 98.51 
      & ~\newline 100.0 \newline 100.0 
      & ~\newline \NR \newline \NR 
      & ~\newline \NR \newline \NR 
      & Signals from retinal birefringence scanning $\rightarrow$ two-layer feed-forward neural net\\
      
      \midrule
      
      \multicolumn{2}{l}{\textbf{Vision screening}}   &  & & & & \\
      
      \tblindent  AVVDA \citep{VanEenwyk2008ArtificialFactors} 
        \newline \tblindent (\citeyear{VanEenwyk2008ArtificialFactors})
      & Strabismus and/or RE   \newline Strabismus \newline High refractive error (RE)
      & \NR \newline 82 \newline 90 
      & \NR \newline \NR \newline \NR
      & \NR \newline \NR \newline \NR 
      & 76.9 \newline \NR \newline \NR
      & Features from Br\"{u}ckner red reflex imaging and eccentric fixation video $\rightarrow$ C4.5 decision tree\\
      
      \midrule
      
      \multicolumn{2}{l}{\textbf{Reading disability (RD)}}   &  & & & & \\
      
      \tblindent  SVM-RFE \citep{NilssonBenfatto2016ScreeningReading} 
        \newline \tblindent (2016)
      & High risk for RD, ages 8--9
      & 95.5 & 95.7 & \NR
      & 95.6
      & SVM with feature selection trained on eye-tracking data\\
      
      \mbox{\tblindent  Polynomial SVM\hspace{0.6ex}\citep{Rello2015DetectingMeasures}} 
        \newline \tblindent (\citeyear{Rello2015DetectingMeasures})
      & \mbox{RD in adults, children ages 11+}
      & \NR & \NR & \NR
      & 80.18
      & SVM trained on eye-tracking and demographic features\\

   \end{tabular}

   \begin{tabular}{@{\kern\tabcolsep}p{1.1in}p{1.5in}p{0.595in}p{0.595in}p{0.595in}p{1.5in}@{\kern\tabcolsep}} 
       \toprule
       \rowcolor{headercolor} \textbf{Approach} \newline \textbf{(Approx.~devel.~year)}  &  \textbf{Predicted category} &   \textbf{AUROC\newline(at 3 years)} & \textbf {AUROC\newline(at 5 years)} & \textbf {AUROC\newline(at 8 years)}& \textbf{Method summary} 
       \global\rownum=1\relax  
       \\
      
      \midrule
      
      \multicolumn{2}{l}{\textbf{Refractive error (RE)}}   &  & & & \\
      
      \tblindent  Random forest  \citep{Lin2018PredictionStudy}
      \newline \tblindent (\citeyear{Lin2018PredictionStudy})
      & Internal evaluation \newline \tblindent High myopia onset \newline Clinical test\newline \tblindent High myopia onset\newline \tblindent High myopia at age 18
      & ~\newline 0.903-0.986 \newline \tblindent \newline 0.874-0.976 \newline     0.940-0.985 
      & ~\newline 0.875-0.901 \newline \tblindent \newline 0.847-0.921 \newline    0.856-0.901 
      & ~\newline 0.852-0.888 \newline \tblindent \newline 0.802-0.886 \newline    0.801-0.837
      & Age, spherical equivalent (SE), and progression rate of SE between two visits was used by a random forest for prediction\\
      
      \bottomrule
    \end{tabular}
    \end{minipage}
    }
    
\end{table*}

\section{Clinical applications of AI}

This section surveys recent AI applications to pediatric ophthalmology, organized by disease (see Table~\ref{tab:ROPsummary}).  The approaches discussed in this survey would more precisely be called applications of ML---the largest subfield of AI concerned with learning models from data.  We have provided a brief overview of AI and ML and their relationship in supplemental material, but the interested reader is encouraged to consult a more extensive tutorial on these topics \citep[e.g.][]{Consejo2019IntroductionOphthalmologists}.   To limit its scope, this review focuses on applications with a goal of having the AI aspects directly impact clinical practice; we omit studies where ML was used primarily for statistical analysis.

\subsection{Retinopathy of Prematurity (ROP)}
The most significant AI advances in pediatric ophthalmology apply to ROP, a leading cause of childhood blindness worldwide \citep{Steinkuller1999ChildhoodBlindness,Gilbert2008RetinopathyControl,Quinn2016RetinopathyEpidemic}. 
In addition to the shortage of trained providers \citep{AmericanAcademyofOphthalmology2006OphthalmologistsCondition,Gilbert2008RetinopathyControl,Quinn2016RetinopathyEpidemic}, ROP exams are difficult, clinical impressions are subjective and vary among examiners \citep{Bolon-Canedoa2015DealingApproach,Wallace2008AgreementPrematurity,Ataer-Cansizoglu2015AnalysisDiagnosis}, and disease management is time-intensive, requiring several serial exams. AI applications have focused on detecting the presence and grading of ROP or plus disease from digital fundus photos. Beyond the benefits of automated ROP screening and objective assessment, digital retinal imaging may cause less pain and stress for infants undergoing ROP screening compared to indirect ophthalmoscopy \citep{Moral-Pumarega2012PainImaging} and enable neonatology-led screening programs \citep{Gilbert2016PotentialTreatment}.


Early computational approaches to detecting plus disease from fundus images focused on vessel tortuosity.  One early attempt to objectively quantify tortuosity used the spatial frequency of manual vessel tracings \citep{Capowski1995APrematurity}.  Since then, there have been several tools developed to determine vessel tortuosity and width via classic image analysis, including Vessel Finder \citep{Heneghan2002CharacterizationAnalysis}, VesselMap \citep{Rabinowitz2007ProgressionAge}, ROPtool \citep{Wallace2007APrematurity}, RISA \citep{Swanson2003SemiautomatedROP,Gelman2007PlusDiagnosis,Gelman2005DiagnosisAnalysis}, CAIAR \citep{Wilson2008ComputerizedInfants,Shah2009SemiautomatedPrematurity}, and IVAN \citep{Shah2009SemiautomatedPrematurity,Sherry2002ReliabilityPopulation}, all of which require at least one manual step from the user.  Recent work suggests other potential vessel measurements correlated with plus disease, such as a decrease in the openness of the major temporal arcade angle  \citep{Oloumi2014QuantificationDisease}.  
Once extracted, retinal vessel measurements have been used as features for various predictive models of plus disease, including linear models such as logistic regression \citep{Gelman2007PlusDiagnosis} and na\"{i}ve Bayes \citep{Bolon-Canedoa2015DealingApproach}, as well as non-linear models trained by support vector machines (SVMs) \citep{Ataer-Cansizoglu2015Computer-basedDiagnosis}.  
 For predicting ROP, \citeauthor{Rani2016MultiplePrematurity} \citep{Rani2016MultiplePrematurity} also employ an SVM, but instead use SIFT \citep{Lowe2004DistinctiveKeypoints} features extracted from retinal image patches and frame the problem in a multiple instance learning \citep{Dietterich2002SolvingRectangles} setting.

Recent approaches to ROP and plus disease detection are mostly based on convolutional neural networks (CNN), which take fundus images as input and do not require manual annotation.   These systems, which include \citeauthor{Worrall2016AutomatedNetworks} \citep{Worrall2016AutomatedNetworks}, i-ROP-DL \citep{Brown2018AutomatedNetworks,Redd2018EvaluationPrematurity}, and DeepROP \citep{Wang2018AutomatedNetworks}, demonstrate agreement with expert opinion \citep{Worrall2016AutomatedNetworks,Redd2018EvaluationPrematurity} and better disease detection than some experts \citep{Brown2018AutomatedNetworks,Wang2018AutomatedNetworks}.

Like many ML methods, these systems can provide a confidence score in their predictions. i-ROP-DL exploits this notion directly by combining the prediction probabilities via a linear formula to compute an ROP severity score, which can serve as an objective quantification of disease; a similar idea could provide finer grading of plus disease \citep{Brown2018AutomatedNetworks}. 

For their core predictive networks, all these CNN-based systems use versions of the Inception architecture \citep{Szegedy2015GoingConvolutions,Ioffe2015BatchShift} with transfer learning \citep{Pan2010ALearning, Weiss2016ALearning} by pretraining on ImageNet, giving them similar foundations.  
However, these approaches differ in preprocessing (e.g., i-ROP-DL \citep{Brown2018AutomatedNetworks} uses a U-net \citep{Ronneberger2015U-net:Segmentation} to perform automatic vessel segmentation) and postprocessing (e.g., i-ROP-DL \citep{Redd2018EvaluationPrematurity} outputs the ROP severity score; \citeauthor{Worrall2016AutomatedNetworks} \citep{Worrall2016AutomatedNetworks} outputs the Bayesian posterior). DeepROP processes a set of fundus images per case, taking a multiple instance learning \citep{Dietterich2002SolvingRectangles} approach, while the other two deep learning methods classify single images.  The other key difference is that these systems are trained on different non-public ROP data sets of varying sizes and labelings (Table~\ref{tab:ROPDLdatasets}).   
The use of non-public data sets and closed implementations (only DeepROP is open source) complicates comparison and reproducibility \citep{Celi2019TheData}.

\setlength{\tblhangindention}{0.75em}

\begin{table}[!tb]
    \centering
    \footnotesize
    
    \renewcommand{\arraystretch}{1.098} 
    \setlength{\tabcolsep}{0.8\tabcolsep}
    \setlength{\aboverulesep}{0ex} 
    \setlength{\belowrulesep}{0ex}
    \sffamily

    \caption{Pediatric ROP data sets used in deep learning}
    \label{tab:ROPDLdatasets}

    \noindent\adjustbox{max width=\textwidth}{
    
    \begin{minipage}{1.0135\textwidth}
    \begin{tabular}{@{\kern\tabcolsep}p{0.6in}p{0.42in}>{\RaggedLeft}p{0.37in}>{\RaggedLeft}p{0.32in}>{\RaggedRight}p{0.85in}@{\kern\tabcolsep}} 
    \toprule
      \rowcolor{headercolor} 
      \textbf{Approach} &
      \mbox{\textbf{Data set}}   & 
      \mbox{\textbf{Patients}} & 
      \textbf{Images} & 
      \textbf{Labels} 
      \\\midrule
      
     DeepROP\newline\citep{Wang2018AutomatedNetworks} 
     & Chengdu
     & 1,273  
     & 20,795
     & \mbox{normal, mild ROP,} severe ROP
     \\

     i-ROP-DL\newline\citep{Brown2018AutomatedNetworks}
     & i-ROP 
     & 898  
     & 5,511
     & normal, plus, \mbox{pre-plus}
     \\

     \mbox{CNN\! +\! Bayes}\newline \citep{Worrall2016AutomatedNetworks} 
     & Canada \newline London
     & \raggedleft\noindent 35 \newline \raggedleft\noindent \NR
     & 1,459\newline 106 
     & normal, plus \newline normal, plus
     \\
            \bottomrule
    \end{tabular}
    \end{minipage}
    }
\end{table}

Current methods for ROP detection are capable of coarse-grained classification, such as discriminating severe from mild ROP; they do not specifically assess disease stage or zone (e.g., \citep{Wang2018AutomatedNetworks}). In fact, all systems except DeepROP \citep{Wang2018AutomatedNetworks} and MiGraph \citep{Rani2016MultiplePrematurity} examine only the posterior pole view, either ignoring other views or explicitly cropping them out. While the literature suggests that severe disease rarely develops without changes in posterior pole vasculature \citep{EarlyTreatmentForRetinopathyOfPrematurityCooperativeGroup2003RevisedTrial}, providing additional outputs of the zone and stage could improve the interpretability of the system's assessment and improve performance.

\subsection{Pediatric Cataracts}
Pediatric cataracts are more variable than adult cataracts, and surgical removal depends upon cataract severity and deprivational amblyopia risk. 
Slit lamp exams enable cataract visualization but can be challenging and subjective, and slit lamp image quality can vary (e.g., based on the child's cooperativeness, image amplification, and interference from eyelashes and other eye disease or structures)  \citep{Liu2017LocalizationNetwork}.  

CC-Cruiser \citep{Long2017AnCataracts,Liu2017LocalizationNetwork,Lin2019DiagnosticTrial} is a cloud-based platform that can automatically detect cataracts from slit-lamp images, grade them, and recommend treatment.  After automatically cropping the slit-lamp image to the lens region, it uses three separate CNNs (modified AlexNets \citep{Krizhevsky2012ImageNetNetworks}) to predict three aspects: cataract presence, grading (opacity area, density, location), and treatment recommendation (surgery or non-surgical follow-up).  CC-Cruiser was evaluated in a multi-center randomized controlled trial within five ophthalmology clinics, demonstrating significantly lower performance in diagnosing cataracts (87.4\%) and recommending treatment (70.8\%) than experts (99.1\% and 96.7\%, respectively), but achieving high patient satisfaction for its rapid evaluation~\citep{Lin2019DiagnosticTrial}.

Children who require surgery face potential complications that differ from those that adults face \citep{Whitman2014ComplicationsSurgery}.  \citeauthor{Zhang2019PredictionMining} applied random forests and na\"{i}ve Bayes classifiers to predict two common post-operative complications, central lens regrowth and high intraocular pressure (IOP),  from a patient's demographic information and cataract severity evaluation  \citep{Zhang2019PredictionMining}. Another approach \citep{Jiang2017AutomaticNetwork} uses a CNN to detect severe posterior capsular opacification warranting surgery, employing a ResNet \citep{He2016DeepRecognition} pretrained on ImageNet with a cost-sensitive loss to handle data set imbalance.

\subsection {Strabismus}
Strabismus affects 1 in 50 children and can cause amblyopia, interfere with binocularity, and have lasting psychosocial effects \citep{Elston1997ConcomitantStrabismus,Adams2003UpdateAmblyopia.,Mojon-Azzi2009StrabismusHeadhunters,Mojon-Azzi2011TheAdults,Mohney2008MentalChildren}. A CNN was used to detect strabismus based on visual manifestation in the eye regions of facial photos \citep{Lu2018AutomatedApplications}, which would be especially useful for telemedical evaluation. For in-office evaluation, which in contrast permits the use of specialized screening instruments, strabismus can be detected using a CNN based on fixation deviations from eye-tracking data \citep{Chen2018StrabismusNetworks}, or with very high sensitivity and specificity from retinal birefringence scanning  \citep{Gramatikov2017DetectingScanning}.

\subsection{Vision Screening}
Like strabismus, refractive error can cause amblyopia, but is difficult for pediatricians to detect. Instrument-based vision screening is recommended \citep{AmericanAcademyofPediatrics2016VisualPediatricians} and most devices have adjustable thresholds for signaling a screening failure. Using video frames from one such instrument that combines Br\"{u}ckner pupil red reflex imaging and eccentric photorefraction, \citeauthor{VanEenwyk2008ArtificialFactors}~trained a variety of ML classifiers to detect amblyogenic risk factors in young children, with the most successful being a C4.5 decision tree \citep{Quinlan1993C4.5:Learning}. 

\subsection{Reading Disability}
Reading disability affects approximately 10\% of children \citep{Rello2015DetectingMeasures}, but objective and efficient testing for it is lacking \citep{NilssonBenfatto2016ScreeningReading}.
Abnormal eye tracking is non-causally associated with reading disability \citep{Rello2015DetectingMeasures,NilssonBenfatto2016ScreeningReading}. 
Two studies used SVMs to identify reading disability from eye movements during reading, either predicting reading disability risk in children ages 8--9 \citep{NilssonBenfatto2016ScreeningReading}, or detecting reading disability in adults and children ages 11+ \citep{Rello2015DetectingMeasures}. The children in both of these studies are older than the optimal age for diagnosis, so validation in a younger cohort could be useful. 

\subsection{Refractive Error}
High myopia is associated with numerous vision-threatening complications \citep{Ikuno2017OverviewMyopia}. Children at risk for high myopia can take low-dose atropine to halt or slow myopic progression \citep{Clark2015AtropineMyopia,Chia2016Five-yearEyedrops}\footnote{Note: this usage of atropine is not approved by the FDA.}, but it can be difficult to determine for which children to recommend this treatment \citep{Lin2018PredictionStudy}. \citeauthor{Lin2018PredictionStudy} \citep{Lin2018PredictionStudy} predicted high myopia in children from clinical measures using a random forest, showing good predictive performance for up to 8 years into the future. Further work has the potential to guide prophylactic treatment. 

\subsection{Non-Pediatric Applications}

AI has been applied to various adult ophthalmic diseases, including diabetic retinopathy \citep{Gargeya2017AutomatedLearning,Soto-Pedre2014EvaluationWorkload,Krause2018GraderRetinopathy,Gulshan2016DevelopmentPhotographs,Pujitha2018RetinalDevelopment}, AMD \citep{Lee2017DeepImages, Rohm2018PredictingDegeneration,Klimscha2017SpatialDegeneration,Bogunovic2017MachineImaging,Grassmann2018APhotography,Schlanitz2017DrusenDegeneration}, sight-threatening retinal disease \citep{DeFauw2018ClinicallyDisease,Ohsugi2017AccuracyDetachment,Zhen2019AssessmentLearning,Schlegl2018FullyLearning,Prahs2017OCT-basedMedications,Bagheri2014EmpiricalData,Kermany2018IdentifyingLearning}, glaucoma \citep{Omodaka2017ClassificationParameters,Li2018EfficacyPhotographs, Martin2018UseGlaucoma}, intraocular lens calculation \citep{Clarke1997ComparisonFormula}, and keratoconus \citep{Hwang2018DistinguishingAnalysis}. It has also been used for robot-assisted repair of epiretinal membranes \citep{Edwards2018First-in-humanSurgery}, retinal vessel segmentation \citep{Lahiri2016DeepAngiography,Maji2015DeepImages,Knudtson2003RevisedDiameters,Ng2010MaximumAnalysis}, and systemic disease prediction from fundus images \citep{Poplin2018PredictionLearning}.  For a detailed review, see \citep{Roach2017ArtificialIntelligence,Consejo2019IntroductionOphthalmologists,Ting2018Artificial,Lee2017MachineArrived,Rahimy2018DeepOphthalmology,Caixinha2017MachineSciences,AmericanAcademyofOphthalmology2018TheOphthalmology, Du2018ApplicationOphthalmology}.

\section{Other ophthalmic applications}

This section reviews applications of ML to pediatric ophthalmology that are not tied to specific diagnoses.

\subsection{Visual Development}

ML has the potential to provide scientific insight into visual development.  For example, adults who had cataract surgery and aphakic correction in infancy have exhibited diminished facial processing capabilities \citep{Lewis2013TheDetection,Grady2014EarlyNetwork}.  This impairment was originally blamed on early visual deprivation \citep{Lewis2013TheDetection,Grady2014EarlyNetwork}, but more recently, it was conjectured to be caused by the aphakic correction and high initial acuity experienced by these infants \citep{Vogelsang2018PotentialAcuity}.  
The hypothesis is that many visual proficiencies, such as facial recognition, are facilitated by the gradual increase in visual acuity during normal visual development.  When tested in CNNs via initial training with blurred images, gradual acuity development increased generalization performance and encouraged the development of receptive fields with a broader spatial extent \citep{Vogelsang2018PotentialAcuity}.  These results provide a possible explanation for the decreased visual proficiencies of congenital cataract patients, and suggest the potential for temporary refractive undercorrection to help restore visual development \citep{Vogelsang2018PotentialAcuity}.

\subsection{Pediatric Retinal Vessel Segmentation}

Although many programs have been developed for vessel segmentation in adults or premature infants, fundus images in older children have unique traits, including light artifacts, that complicate segmentation \citep{Fraz2014DelineationClassification}.   \citeauthor{Fraz2014DelineationClassification} \citep{Fraz2014DelineationClassification} developed an ensemble of bagged decision trees that use multi-scale analysis with multiple filter types to do vessel segmentation in pediatric fundus images. Another tool, CAIAR \citep{Wilson2008ComputerizedInfants}, has been validated in school-aged children \citep{Owen2009MeasuringProgram}. CAIAR was first applied to infants with ROP and uses a generative model of the vessels fit via maximum likelihood to a multi-scale representation of the retinal image \citep{Wilson2008ComputerizedInfants}.

\subsection{Ophthalmic Image Synthesis}

Through their multi-layered representation, deep learning methods such as generative adversarial networks \citep{Goodfellow2014GenerativeNets} are able to synthesize novel realistic images, including retinal fundus images \citep{Zhao2017SynthesizingGANs,Costa2018End-to-endSynthesis}. Such synthesized images can compensate for data scarcity, preserve patient privacy, and depict variations on or combinations of diseases for resident \mbox{education \citep{Yi2019GenerativeReview,Finlayson2018TowardsEducation}.}  

\begin{figure*}[!t]
    \centering
    \includegraphics[width=0.8\linewidth]{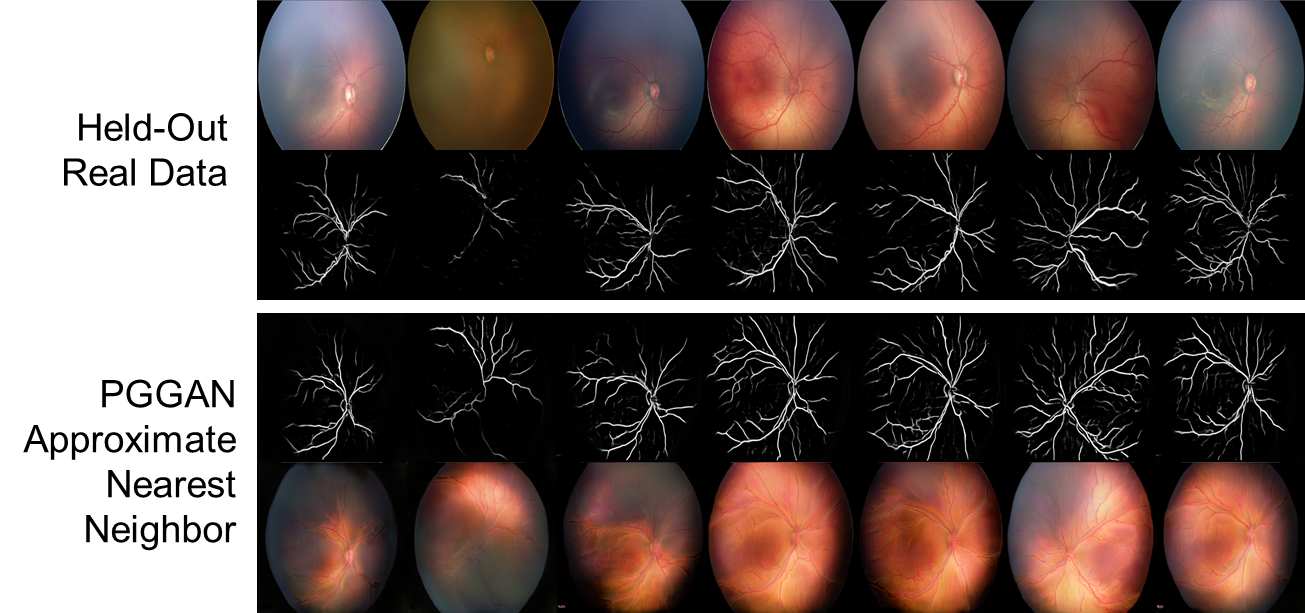}
    \caption{Real (top row) and synthetic (bottom row) fundus images of ROP with their corresponding vessel segmentations \citep{Beers2018High-resolutionNetworks}. The top row shows real images that were not included in the training set, and the bottom row shows the most similar synthesized images. (Image from \citep{Beers2018High-resolutionNetworks}, reused with permission.)}
    \label{fig:Beers2018HighResolution-SyntheticImages}
\end{figure*}

One recent technique to synthesize high-resolution images, progressive growing of GANs (PGGANs), was used to synthesize realistic fundus images of ROP (see examples in Figure~\ref{fig:Beers2018HighResolution-SyntheticImages}) \citep{Beers2018High-resolutionNetworks}.  The PGGAN was trained on ROP fundus images in combination with vessel segmentation maps obtained from a pre-trained U-net CNN \citep{Ronneberger2015U-net:Segmentation}. GANs have also been used to synthesize retinal images of diabetic retinopathy, including the ability to control high-level aspects of the presentation \citep{Pujitha2018RetinalDevelopment,Niu2019PathologicalDiagnosis}.  While many of the GAN-synthesized images display believable pathologic features,  
some do contain ``checkerboard'' and other generation artifacts. 

\section{Current Limitations and\\Future Directions}

Current applications to pediatric ophthalmology have several limitations that offer avenues for future work.

\paragraph{Disagreement on reference standards}  
An ML classifier's performance is fundamentally limited by the quality of the training data, which are manually labeled by clinicians.  However, there is often significant variation of the diagnosis and treatment among physicians, given the same case information \citep{Chiang2007InterexpertPrematurity,Bolon-Canedoa2015DealingApproach,Ataer-Cansizoglu2015AnalysisDiagnosis, Wallace2008AgreementPrematurity}, which complicates determination of the correct labels. When ML was used to identify factors influencing ROP experts' decisions for plus disease diagnosis, the most important features were venous tortuosity and vascular branching \citep{Bolon-Canedoa2015DealingApproach,Ataer-Cansizoglu2015AnalysisDiagnosis}, neither of which are part of the standard ``plus disease'' definition of arteriolar tortuosity and venular dilatation \citep{CommitteefortheClassificationofRetinopathyofPrematurity1984AnPrematurity,InternationalCommitteefortheClassificationofRetinopathyofPrematurity2005TheRevisited}. Most approaches use the majority label from multiple experts as the label for each training instance, or combine the majority label given to imagery with the clinical diagnosis \citep{Ryan2014DevelopmentOphthalmology}. An alternative approach puts cases with any amount of disagreement up for adjudication among the experts, resulting in a consensus label and reducing errors, as demonstrated for  diabetic retinopathy \citep{Krause2018GraderRetinopathy}.

\paragraph{Need for pediatric-specific models}

It would be advantageous for pediatric ophthalmology to benefit from the large amount of work in AI for adult ophthalmology.  However, due to the unique aspects of pediatric disease manifestation, ML models trained on adult patients may make errors when directly applied to pediatric patients.  Transfer learning  \citep{Pan2010ALearning,Weiss2016ALearning} and  multi-task learning \citep{Ruder2017AnNetworks,Zhang2018AnLearning} techniques may offer a solution to this problem, providing mechanisms to adapt adult models to pediatric patients given a small amount of pediatric ophthalmic data.  
These methods could also reuse knowledge across models of different diseases or populations---for example, integrating knowledge across multiple smaller pediatric data sets of different ophthalmic diseases to help compensate for the lack of data on any one disease.  Notice that, by pretraining on ImageNet, many of the CNN-based methods surveyed here already employ transfer learning of basic image features to compensate for using small data sets; transferring from adult ophthalmic data sets may provide further advantages.

\paragraph{Poor reproducibility and comparability}

Almost all the ML studies discussed here, even those that focus on the same disease, are trained and evaluated on different data sets.  In many cases, the data sets and software source code are not available publicly, complicating reproducibility and scientific comparison across algorithms \citep{Celi2019TheData}.

Most ML research relies on publicly accessible data sets and software implementations for evaluation and comparison.
One simple way to encourage further applications of AI to pediatric ophthalmology is through the public release of data sets in strict compliance with HIPAA regulations, and with special regard to the additional HIPAA restrictions for minors. 
Even small pediatric ophthalmic data sets could be of use
when used in combination with adult data through transfer learning techniques, as mentioned above.  For the largest impact, these open data sets should be hosted in a widely used ML repository.

\paragraph{Lack of temporal information}  
Most of these systems detect disease based upon one snapshot in time, without consideration of longitudinal imaging of the case  \citep{Worrall2016AutomatedNetworks}. In some diseases, such as ROP, rapid change is associated with poorer outcomes \citep{Heneghan2002CharacterizationAnalysis,Wallace2000PrognosticPrematurity}, suggesting that temporal information may have a role in predicting severe disease.

\paragraph{Uninterpretable ``black-box'' models}

Despite their predictive power, the ``black-box'' nature of most state-of-the-art ML methods, such as deep neural networks, complicates their application in medicine.  It is often challenging to quantitatively interpret the inference process of such models, understanding how they arrived at their predictions \citep{Doshi-Velez2017TowardsLearning,Gilpin2018ExplainingLearning}. Since they focus on correlations between the input and desired output, in some cases ML models may fixate on confounding factors instead of pathological information \citep{Zech2018VariableStudy}. Interpretable ML methods provide a potential solution to benefit clinicians, allowing, for example, examination of intermediate decision steps within a deep network, natural language justifications for a decision, or visualization of image features that contribute to a decision \citep{Gilpin2018ExplainingLearning}.  While these methods seek to improve the interpretability of black-box models, other approaches seek to improve the predictive power of models that are already interpretable, such as the MediBoost algorithm for growing decision trees via gradient boosting \citep{Valdes2016MediBoost:Medicine}.

\section{Conclusion}
There is a large potential for current and future AI applications to pediatric ophthalmology, and there are some diseases, such as NLDO, congenital glaucoma, and congenital ptosis, without any published applications of AI to our knowledge. Automated disease detection, the most common use case, could augment telemedical efforts to broaden access to care, improve efficiency, and result in earlier diagnoses. However, other less-utilized capabilities of this technology, including disease grading and outcome prediction, have the potential to enhance clinical care. All AI methods deployed in clinical care must ultimately match or surpass physician performance while meeting the unique requirements of both clinicians and pediatric patients, suggesting the need to augment evaluations on experimental data sets with clinical~trials.


\subsection{Acknowledgements} 

\textit{We would like to thank Jing Jin, MD, Jos\'{e} Marcio Luna, PhD, and Jorge Mendez for their helpful feedback on  this article.}


\subsection{Financial support and sponsorship} 

\textit{E.E.'s work was partially supported by the Lifelong Learning Machines program from DARPA/MTO under grant \#FA8750-18-2-0117. The funders had no role in the research presented in this article, nor in its preparation, review, or approval. The views and conclusions contained herein are those of the authors and should not be interpreted as necessarily representing the official policies or endorsements, either expressed or implied, of DARPA or the U.S.~Government.}


\subsection{Conflicts of interest}

\textit{There are no conflicts of interest.}


\defbibnote{myprenote}{{\sffamily\selectfont Papers of particular interest, published within the annual period of review, have been highlighted as:\\\emphrefbib\phantom{\emphrefbib}\hspace{1em} of special interest\\\emphrefbib\emphrefbib\hspace{1em} of outstanding interest}}
\printbibliography[prenote=myprenote]

\clearpage

\appendix

\section{Online Supplement:\\A Brief Overview of AI and ML}


Artificial intelligence (AI) is the broad field concerned with the study of intelligence and its computational manifestation within machines.  It spans a broad set of problems that are all interrelated, from basic search (e.g., route finding on a map, or sequences of moves in a chess game) to logical reasoning (e.g., theorem proving, logistics planning) to reasoning under uncertainty (e.g., Bayesian abductive reasoning) to multi-agent systems (e.g., markets of trading agents) to robotics (e.g., computer vision and perception, control of dynamical systems) to learning.

Machine learning (ML) is perhaps the largest subfield of AI and is concerned with this latter problem of learning from experience (i.e., data).  Most recent news headlines and research concerning the application of AI techniques to problems in other disciplines (including the title of this article) would more precisely be termed applications of ML.  Modern statistical ML is primarily concerned with the optimization of a model (e.g., a classification or regression model) to fit a given set of training data in such a manner that the model will be able to generalize to new data.  

As a simple example, the training data might consist of demographical, biometric, and imaging data of a chosen cohort of 10,000 patients gathered from hospital records.  Each record (called a {\em data instance}) could be characterized as a set of categorical, ordinal, and numeric features that are derived from the patient's record, and would be labeled according to the patient's diagnosis.  ML algorithms could then train a classifier model (e.g., a decision tree, logistic regression) to predict the labeled diagnosis of a patient given the set of features derived from their record.  Critically, the performance of the classifier should be assessed on new patients from the same population (i.e., patients with similar demographics that were not present in the training data), using application-dependent metrics (such as accuracy, sensitivity/specificity, receiver operating characteristic (ROC) curves).  

This example focused on a {\em supervised learning} setting, in which each patient's data instance had a corresponding categorical label and we trained a classification model.  If the labels had instead been numeric values, we could have trained a regression model using other supervised learning algorithms.  Other settings include {\em semi-supervised learning}, in which only some data instances are labeled; {\em unsupervised learning}, which focuses on discovering patterns in unlabeled data (e.g, clusters of patients with similar biomarkers), and {\em reinforcement learning}, which seeks to learn a policy that can determine sequences of actions to execute to achieve a goal (e.g., the sequence of treatments to administer to an ICU patient, or the movements a robot should perform to tie a ligature).  There are numerous different ML techniques, which vary according to the model representation (e.g.,  decision trees, linear classifiers, neural networks, logical rules), the mathematical technique used to optimize the model (e.g., greedy heuristics, gradient descent, evolutionary computation), and the evaluation metric used to assess the quality of model fit to the data (e.g., accuracy, precision and recall, posterior probability). 
Note that these metrics focus on performance on data and do not necessarily relate to the model acquiring generalizable knowledge.   Consequently, ML models are learning patterns of correlations between the inputs and the desired outputs, not causal knowledge.  This may cause them to exploit confounding details instead of physiological aspects. For example, an image classifier tasked with predicting disease severity might erroneously focus on identifying the type of camera (portable vs. fixed) or the presence of chest tubes or other medical devices, rather than pathological information, simply because these other confounding details are highly correlated with the desired output \citep{Zech2018VariableStudy}. 

Deep learning (DL) methods are one subgroup of ML techniques that have shown exceptional impact to a wide variety of applications.  Although DL techniques have been studied for decades, recent advances in computational algorithms and hardware have enabled these models to be trained at scale on large data sets, leading to their impact.  DL is concerned with training models with numerous layers of processing, such as deep neural networks.  Convolutional neural networks (CNNs) are one popular type of deep network that are often used for image classification.  These models take raw input, such as a fundus photograph, and extract layered features from the input image, where higher levels of the deep neural network typically focus on increasingly abstract features that are built upon lower-level features.  This automatic discovery of features is called representation learning, since the model identifies commonalities within the given input data as a way to re-represent it at different levels of abstraction.  Although fundamentally an unsupervised learning technique, deep learning models can easily be adapted for classification, regression, and reinforcement learning. 
Despite its success and popularity, DL typically requires large data sets for training (e.g., thousands or hundreds of thousands or millions of examples, depending on the complexity of the decision), which may be problematic in certain medical applications. 

\end{document}